\documentclass[twocolumn]{aastex63}

\definecolor{amethyst}{rgb}{0.6, 0.4, 0.8}

\submitjournal{APJL}

\shorttitle{ALMA long-baseline observations of ISO-Oph 2}
\shortauthors{Gonzalez-Ruilova et al.}

\begin{document}

\title{A Tale of Two Transition Disks: ALMA long-baseline observations of ISO-Oph 2 
reveal two closely packed non-axisymmetric rings and a $\sim$2 au cavity
\footnote{Released on June, 10th, 2019}}

\author[0000-0003-4907-189X]{Camilo González-Ruilova}
\affiliation{Facultad de Ingeniería y Ciencias, Núcleo de Astronomía, Universidad Diego Portales, Av. Ejercito 441, Santiago, Chile}

\author[0000-0002-2828-1153]{Lucas A. Cieza}
\affiliation{Facultad de Ingeniería y Ciencias, Núcleo de Astronomía, Universidad Diego Portales, Av. Ejercito 441, Santiago, Chile}

\author[0000-0001-5073-2849]{Antonio S. Hales}
\affiliation{Joint ALMA Observatory, Alonso de Cordova 3107, Vitacura 763-0355, Santiago, Chile}
\affiliation{National Radio Astronomy Observatory, 520 Edgemont Road, Charlottesville, VA 22903-2475, USA}

\author[0000-0003-2953-755X]{Sebasti\'an P\'erez}
\affiliation{Departamento de F\'isica, Universidad de Santiago de Chile, Av. Ecuador 3493, Estaci\'on Central, Santiago, Chile}

\author[0000-0002-5903-8316]{Alice Zurlo}
\affiliation{Facultad de Ingeniería y Ciencias, Núcleo de Astronomía, Universidad Diego Portales, Av. Ejercito 441, Santiago, Chile}
\affiliation{Escuela de Ingenier\'ia Industrial, Facultad de Ingenier\'ia y Ciencias, Universidad Diego Portales, Av. Ejercito 441, Santiago, Chile}
%%%%%%%%%%%%%%%%%%%%%%%%%%%%

\author[0000-0002-0176-4331]{Carla Arce-Tord}
\affiliation{Universidad de Chile, Camino el Observatorio 1515, Santiago, Chile}

\author{Simón Casassus}
\affiliation{Universidad de Chile, Camino el Observatorio 1515, Santiago, Chile}

\author[0000-0001-7668-8022]{Hector C\'anovas}
\affiliation{Aurora Technology B.V. for ESA, ESA-ESAC, Camino Bajo del Castillo s/n, E-28692 Villanueva de la Ca\~nada, Madrid, Spain}

\author{Mario Flock}
\affiliation{Max Planck Institute for Astronomy, Königstuhl 17, 69117 Heidelberg, Germany}

\author{Gregory J. Herczeg}
\affiliation{Kavli Institute for Astronomy and Astrophysics, Peking University, Yiheyuan Lu 5, Haidian Qu, 100871 Beijing, People's Republic of China)}

\author{Paola Pinilla}
\affiliation{Max Planck Institute for Astronomy, Königstuhl 17, 69117 Heidelberg, Germany}

\author[0000-0002-4716-4235]{Daniel J. Price}
\affiliation{School of Physics \& Astronomy, Monash University, VIC 3800, Australia}

\author[0000-0002-7939-377X]{David A. Principe}
\affiliation{MIT Kavli Institute for Astrophysics and Space Research, Cambridge, MA, USA}

\author[0000-0003-3573-8163]{Dary Ru\'iz-Rodr\'iguez}
\affiliation{National Radio Astronomy Observatory, 520 Edgemont Road, Charlottesville, VA 22903-2475, USA}

\author[0000-0001-5058-695X]{Jonathan P. Williams}
\affiliation{Institute for Astronomy, University of Hawaii at Manoa, Honolulu, HI, 96822, USA}

\begin{abstract}
ISO-Oph 2 is a wide-separation (240 au) binary system where the primary star harbors a massive (M$_{\rm dust}$ $\sim$40 M$_{\oplus}$) ring-like disk with a dust cavity $\sim$50 au in radius and the secondary hosts a much lighter (M$_{\rm dust}$ $\sim$0.8 M$_{\oplus}$) disk.
As part of the high-resolution follow-up of the ``Ophiuchus Disk Survey Employing ALMA” (ODISEA) project, 
we present 1.3 mm continuum and $^{12}$CO molecular line observations of the system at 0\farcs02 (3 au) resolution. 
We resolve the disk around the primary into two non-axisymmetric rings and find that the disk around the secondary is only $\sim$7 au across and also has a dust cavity (r $\sim$2.2 au). 
Based on the infrared flux ratio of the system and the M0 spectral type of the primary, we estimate the mass of the companion to be close to the brown dwarf limit.
Hence, we conclude that the ISO-Oph 2 system contains the largest and smallest cavities, the smallest measured disk size, and the resolved cavity around the lowest mass object (M$_{\star}$ $\sim$0.08 M$_\odot$) in Ophiuchus. 
From the $^{12}$CO data, we find a bridge of gas connecting both disks.
While the morphology of the rings around the primary might be due to an unseen disturber within the cavity,  
we speculate that the bridge might indicate an alternative scenario in which the secondary has recently  flown by the primary star causing the azimuthal asymmetries in its disk. %
The ISO-Oph 2 system is therefore a remarkable laboratory to study disk evolution, planet formation, and companion-disk interactions. 
\end{abstract}


\section{Introduction} \label{sec:intro}

The Atacama Large Millimeter/submillimeter Array (ALMA) is revolutionizing the field of disk evolution and planet formation by studying large populations of protoplanetary disks in nearby molecular clouds and individual objects in unprecedented detail. On one hand, disk demographic surveys provide information on fundamental disk properties such as mass and size as a function of stellar mass \citep{Pascucci2016,Barenfeld2016}, age \citep{Ansdell2017,Ruiz2018}, and multiplicity \citep{Cox2017,Manara2019,Zurlo2020}. On the other hand,  detailed high-resolution studies can deliver images at 3-4 au resolution at the distance of nearby star-forming regions (140--200 pc) such as Taurus, Lupus, Ophiuchus, and Chamaeleon in order to study their substructures \citep{ALMA2015,Andrews2018}.

With almost 300 targets, the ``Ophiuchus Disk Survey Employing ALMA'' (ODISEA) project \citep{Cieza2019,Williams2019} is so far the largest of the disk demography surveys in nearby clouds and has identified several interesting objects for high-resolution follow-up observations. 
One of the most interesting targets identified by ODISEA is ISO-Oph 2, a binary system with a projected separation of 240 au and a flux ratio of 0.08 in the K-band \citep{Ratzka2005}. 
The primary star  has an M0 spectral type and an accretion rate typical of Classical T Tauri stars (10$^{-8.7}$ M$_{\odot}$yr$^{-1}$; \citep{Gatti2006} and hosts a massive (M$_{\rm dust}$ $\sim$40 M$_{\oplus}$) disk with the largest dust cavity ($\sim$50 au in radius) seen in the Ophiuchus cloud \citep{Cieza2019}. The secondary star, which lacks a spectral classification in the literature, arbors  a much lighter (M$_{\rm dust}$ $\sim$ 0.8 M$_{\oplus}$) disk that remained unresolved at 30 au resolution \citep{Cieza2019}.

ISO-Oph 2 was observed with ALMA as part of the high-resolution follow-up of the 10 brightest ODISEA targets that were not included in the ALMA Cycle-4 Large Program DSHARP \citep{Andrews2018}. The full sample of ODISEA targets observed at high-resolution is discussed in Cieza et al. (in prep.). Here we present the initial results of this long-baseline program and show that the ISO-Oph 2 system displays several exceptional properties when observed at 0.02$''$ (3 au) resolution. 

\section{Observations and data analysis} \label{sec:observations}

\subsection{ALMA observations}

The long-baseline data of ISO-Oph 2 were obtained during ALMA Cycle~6 under program 2018.1.00028.S. The source was observed in Band-6 (1.3 mm/230 GHz) in two different epochs with baselines up to 16.2 km. The correlator setup was chosen to maximize the continuum bandwidth (7.5 GHz). Three spectral windows were configured in Time Division Mode, with spectral resolution of 43 ~km~s$^{-1}$, and overlap the continuum frequency from the Cycle~4 ODISEA observations at $\sim$217, 219, and 233~GHz. A fourth baseband was centered in the $^{12}$CO J = 2-1 line (230.538 GHz) and delivers a modest resolution of 1.5 km~s$^{-1}$, enough to isolate the line (line width $\sim$5 ~km~s$^{-1}$) and trace the spatial distribution of the gas. The observing log of the Cycle-6 long-baseline observations are shown in Table~1, together with relevant information on Cycle~4 data at lower spatial resolutions (\citealt{Cieza2019}, PID = 2016.1.00545.S) that we use to complement our new $^{12}$CO observations (see Section 3.3). 

\begin{deluxetable*}{lcccccccccc}
\label{tableobs}
\tablewidth{600pt}
\tabletypesize{\scriptsize}
\tablehead{
\colhead{UTC Date} & 
\colhead{2019-06-21/02:07 } & 
\colhead{2019-06-12/02:15} & 
\colhead{2018-05-26/06:44} & 
\colhead{2017-07-13/03:29}
}
\startdata
Config &C43-8/9 &C43-8/9 &C40-2 &C40-5\\
Baselines &82 m - 16.2 km &82 m - 16.2 km &14 m - 313 m & 16 m - 2.6 km\\
Mean PWV (mm) &0.85 &1.19 &1.07 &1.8 \\
Calibrators &J1517-2422, J1633-2557, &J1517-2422, J1633-2557, &J1517-2422, Ganymede, &J1517-2422, J1517-2422, \\
&J1642-2849&J1642-2849&J1625-2527&J1625-2527\\
Beam size ($''$)$^a$& 0.02 $\times$ 0.03  &  0.02 $\times$ 0.03    &  1.1 $\times$ 1.1 & 0.2 $\times$ 0.3 \\
Spectral resolution (km s$^{-1}$) & 1.5 &  1.5 &0.08& 0.08&\\
Continuum rms (mJy beam$^{-1}$)&3.6$\times$10$^{-2}$&2.7$\times$10$^{-2}$&2.4$\times$10$^{-1}$ & 3.7$\times$10$^{-1}$\\
Line rms (mJy beam$^{-1}$ km s$^{-1}$)& 26.5& 18.7 & 89.5 & 53.4 \\
\tablecaption{ALMA observations of the ISO-Oph 2 system}
\enddata
\tablenotetext{a}{beam size corresponds to Briggs weighting and robust parameter of 0.5.}
\end{deluxetable*}

\subsection{Data analysis}

All data sets were calibrated using the ALMA Science Pipeline which includes  Water Vapor Radiometer and system temperature correction, as well as bandpass, phase, and amplitude calibration.  
 
Continuum and line imaging was performed using the {\sc TCLEAN} task in CASA v.5.6.1  \citep{McMullin2007}, with Briggs weighting and robust parameter of 0.5. 
Continuum images were produced using the continuum spectral windows of the high resolution data only, yielding an image with synthesized beam of  0.02$''$ $\times$ 0.03$''$ centered at 225 GHz.
Since there are two sources in the field of view, manual masks around each source were defined during the CLEANing process. Two iterations of phase-only self-calibration were conducted on each epoch separately using a mask covering both sources.  Self calibration improved the signal-to-noise ratio of the final image by a factor of 1.7. The resulting image is shown in Figure 1, and its properties are described in Table 1.

Line imaging of the $^{12}$CO gas was performed combining the datasets with 0.02$'', $0.2$''$, and 1.1$''$ resolution.
The datasets at 0.02$''$ and 0.2$''$ resolution (2019 and 2017 epochs) were recentered using the location of the  secondary disc at 0.02$''$ resolution in the continuum as reference point.  The  coordinates of the 1.1$''$ images (2018 epoch) were taken at face value as the secondary disc is only barely detected. 
Self-calibration of the line data from all epochs and resolutions was done by applying the self-calibration solutions derived from the respective continuum datasets. We used the  {\sc UVCONTSUB} task to substract the continuum from the line data, after which the data was CLEANed using an auto-masking process (for regions up to 3$\times$rms) in {\sc TCLEAN}  to generate line cubes with 1.5
km~s$^{-1}$ resolution. 
To obtain an optimal balance between spatial resolution and sensitivity, we applied an {\sc uvtaper} with an {\sc uppertaper = 0.1arcsec.}

\section{Results} \label{sec:results}

\subsection{(Sub)stellar masses}

In order to estimate the mass of each component in the binary system, we first use the spectral type of the primary star (M0, \citealt{Gatti2006}) and the infrared photometry to estimate its extinction, mass and age, and then use the flux ratio in the K-band  to estimate the mass of the secondary object assuming the same age and extinction.  We assume that the observed J-K color from 2MASS \citep{Cutri2003} is photospheric and calculate the extinction in K-band as $A_{K} = 0.53 \times    \left[\left(J -K\right)
-\left(J-K\right)_O\right]$, 
where $\left(J-K\right)_O$ is the photospheric color of an M0 star.
This assumption is supported by the fact that the ALMA data place an upper limit of 0.1 M$_{\oplus}$ for the mass of an inner disk around ISO-Oph 2A (see Section~\ref{s:continuum}), but we note that the presence of K-band excess can not be ruled out for neither the A or B component in the system. 
This approach gives  A$_K$ = 1.3 mag and  A$_V$ = 14.5 mag. Adopting the BT-Settl models \citep{Allard2012} and a distance of 144 pc  \citep{Gaia2018}, we estimate that the primary is a $\sim$1 Myr old star with a mass of 0.5 M$_\odot$ based on its spectral type and extinction corrected absolute K-band magnitude, 2.45.
Similarly, the flux ratio indicates that the secondary has a extinction-corrected K-band absolute magnitude of 5.2, corresponding to a T$_{eff}$ of 2900 K and a mass of 0.08 M$_\odot$, right at the stellar/substellar boundary. However, we note that stellar masses and ages of low-mass stars and brown dwarfs are model-dependent and notoriously difficult to  estimate for individual objects \citep{Gonzales2020},
and hence that these values must be taken with caution. 

\subsection{Continuum}\label{s:continuum}

\subsubsection{Primary disk}

The disk around the primary star  shows a large cavity devoid of millimeter emission and a narrow and structured continuum outer disk (see Fig. 1). 
The outer region is bright and breaks into substructure in the form of two narrow  rings, which show spectacular azimuthal asymmetries. %
The outermost structure peaks toward South of the star and presents a prominent decrement to the North-West. 
On the other hand, the innermost structure peaks toward the North-West and becomes nearly undetectable toward the South-East. 
We do not detect emission near the center of the cavity, placing strong constraints on the mass of a putative inner disk.  Based on the 5-$\sigma$ noise (0.18 mJy/beam) of the continuum observations and adopting the same basic assumptions as in \citet{Cieza2019}, this limit corresponds to a dust mass of 0.1 M$_{\oplus}$ for an unresolved inner disk. 

In order to highlight the substructure in the disk, a process of unsharp-masking was applied \citep{Perez2020A}, which accentuates the narrow rings.
This sharper image is obtained by convolving the image with a circular Gaussian kernel to produce a smoothed version of the disk which is later subtracted from the original image.

In Fig. 2, we show the unsharp image of the disk around the primary deprojected to a Position Angle ($PA$) and an inclination ($i$) of 0.0 deg (panel a) and the corresponding radial profile (panel b). 
To perform the deprojection, we explored a grid of different $PA$,  $i$,  and centers. We find that a $PA$ = 1.0 deg (E of N), and an $i$ = 36.4 deg, together with the coordinates from the Gaia Data Release 2 \citep{Gaia2018} corrected for proper motions\footnote{Ra = 246.408838 deg; Dec = $-$24.376830 deg; Epoch = 2015.5; pmra = $-$5.4 mas yr$^{-1}$; pmdec = $-$25.2 mas yr$^{-1}$}  provide the most circular deprojected image.  
This suggests that the primary star is likely to be located at the center of the disk to within  $\sim$15 mas or  2 au).
Both rings show azimuthal variations in intensity and/or width,
which are difficult to disentangle, but as shown in panel b, when averaged over azimuth, the inner ring extends from a radius of $\sim$0.3$''$ to $\sim$0.4$''$ (43 to 58 au), while the outer one does from a radius of $\sim$0.4$''$ to $\sim$0.6$''$ (58 to 86 au).

Fig. 2 (panel b) additionally includes the K-band 5-$\sigma$ contrast as a function of radius (dashed-line) corresponding to the VLT-NACO image of  ISO-Oph 2 presented in Fig.~6 of \cite{Zurlo2020}. 
While the 2.2~$\mu$m NACO data is not particularly deep, it allow us to rule out the presence of an equal brightness binary at the resolution limit of the observations 0.05$''$ (7 au) and of a 0.08 M$_{\odot}$ brown dwarf (K-band contrast $\sim$0.08) at distances larger than 50 au from the primary.
In Fig. 2, we also show the deprojected profile in polar coordinates (panel c), which we also average over the radial range of each ring, 0.3$''$  -- 0.4$''$ for the inner ring and 0.4$''$ -- 0.6$''$ for the outer ring (panel d). 
These radially-averaged azimuthal profiles allow us to better characterize the azimuthal asymmetries. 
The outer disk has a minimum at a $PA$ of $\sim$50 deg West of 
North, but emission is detected at all azimuths. The inner ring is fainter and radially narrower. It has an azimuthal extension of $\sim$260 deg (from $PA$ = 100 deg East of North to 160 deg West of North), peaking close to the minimum seen in the outer disk.  

\subsubsection{Secondary disk}

The disk around the secondary star is only 1.3 mJy and remained unresolved in our 0.2$''$ resolution observations \citep{Cieza2019}.The long-baseline data shows that the disk is very compact, 0.05$''$ (7 au) 
across (FWHM measured with the {\sc IMFIT} routine within {\sc CASA}) and exhibits the double peak (Fig. 1, panel b) characteristic of a transition disk with a marginally resolved inner dust cavity. 

Given the small flux and angular size of the disk, which push the limit of ALMA's capabilities, we estimate the size of the dust cavity adopting an analytical approach based on deprojecting the visibilities to 0.0 deg $i$ and 0.0 deg $PA$ in the $u-v$ plane. 
In particular, the Real component of the visibilities as a function of deprojected $u-v$ distance for a face-on narrow ring is a zeroth-order Bessel function and the imaginary flux components of such a ring is 0.0 at all spatial frequencies. Following \citet{Hughes2007}, the location of the first null on the Bessel function,  $R_{null}$, is related to the cavity radius, $R_{cav}$, as follows:

\begin{equation}
R_{null} (k\lambda) = \frac{77916}{\pi^2} \left(
\frac{Distance}{100~pc}\right)\left(\frac{1~au}{R_{cav}}\right) 
\end{equation}

The deprojected  \textit{u-v}-distances are given by
$R=\sqrt{d^2_a+d_b^2}$, where $d_a =  \sqrt{u^2+v^2}\,  \sin\,  \phi$
and $d_b =  \sqrt{u^2+v^2}\,  \cos\,  \phi \,  \cos \, i$,
$\phi = \arctan(v/u) - PA$ \citep{Lay1997}.

In order to construct the visibility profile of the disk around the secondary star, we first subtracted the visibilities of the CLEAN model of the disk around the primary  from  the  visibility  data, following \citet{Perez2020B}. 
We also changed the phase center to the position of the secondary as measured with {\sc IMFIT} in {\sc CASA}, which also  
provided initial approximations for the $i$ and $PA$ of the secondary disk. 
More precise values were obtained by minimizing the imaginary components in the imaginary part of the visibility profile. The resulting visibility profile corresponding to an $i$ = 10 deg and $PA$ = 122 deg is shown in Fig. 3.  A null is seen at 5000 $\pm$ 500 k$\lambda$, which according to Equation 1 corresponds to cavity radius of 2.2$\pm$ 0.25 au.
Given the small size of the outer disk (FWHM/2 = 3.5 au), we conclude that the secondary disk also shows a ring-like morphology like the primary (e.g.,  with a width smaller than its inner radius).

\twocolumngrid
\subsection{Molecular gas ($^{12}$CO) data}

We find that the molecular line data of ISO-Oph 2 system exhibits interesting features as well. 
The $^{12}$CO line peaks  inside the cavity, but the emission is shifted toward the south, reaching the secondary (Fig. 4, panel a). As shown in the Position-Velocity (P-V) diagram (panel b), the disk around the primary is mostly detected at $\sim$1 and $\sim$4 km/s,  while there is significant cloud contamination at velocities between 2 and 3 km/s. 
The secondary, on the other hand, is mostly detected at $\sim$1.5 km/s. At velocities between 0 and 2 km/s, the P-V diagram also shows what seems to be a bridge of gas connecting both disks. 
This bridge becomes even more evident in the channel map containing emission between 0.25 and 1.75 km/s (panel c). 
%} 
%

Finally, in panel d of Fig. 4, we show the 
intensity-weighted mean velocity (moment-1) map of the $^{12}$CO data. This map is consistent with Keplerian motion as the extreme velocities are observed along the major axis of the disk.  We therefore conclude that there is no indication of an inner cavity in the gas, which is consistent with the relatively high accretion rate of the system  (10$^{-8.7}$ M$_{\odot}$yr$^{-1}$; \citealt{Gatti2006}).
We also note that the kinematic center of the disk (where the minimum and maximum velocities converge) coincides with the expected location of the star based on the position and proper motions from Gaia.

\section{Discussion}\label{sec:conclusion}

ALMA observations at 3-30 au resolution have shown that protoplanetary disks in general, and transition disks in particular, have diverse substructures. 
The DSHARP project, the  largest long-baseline survey of disks thus far \citep{Andrews2018} has revealed that narrow rings and gaps are the most common substructure in protoplanetary disks. Unfortunately that project does not include transition objects. More recently, \citet{Francis2020} collected a large sample of $\sim$40 of disks with large (r $>$ 25 au) dust cavities. 
Around half of the stars in that sample have dusty inner disks detected, something that is not seen in ISO-Oph 2 A. 

Most transition disks are azimuthally symmetric, but several examples  with azimuthal asymmetries do exist such as IRS-Oph~48 \citep{vanderMarel2013}, HD~142527 \citep{Casassus2013}, HD~135344B \citep{Cazzoletti2018}, SR~21 \citep{Muro2020}, CIDA~9, RY~Lup and HD~34282 \citep{Long2019,Francis2020}. 
These asymmetries are most likely dynamical in nature and can be planet-induced vortices \citep{Hammer2019.482.3609H} or be produced by stellar companions \citep{Price2018,Calcino2019}. 
However, the disk around ISO-Oph 2 A is particularly rare because it shows not one, but two closely packed and azimuthally asymmetric rings, resembling other very complex systems such as MWC 758 \citep{2018ApJ...860..124D}.
As demonstrated by \citet{Perez2019}, a single migrating planet can produce multiple rings, but it remains to be established by future hydrodynamic modeling if the rings produces this way can be as non-axisymmetric as seen in ISO-Oph~2~A. 

The possible bridge of gas connecting both disks in the system (Fig. 4) might indicate a different origin for the complex structure of the disk around the primary. At its current location, the secondary itself is too far away (240 au) to disturb the primary disc significantly. 
However, if ISO-Oph 2 is a very eccentric binary system or an unbound flyby, it is possible that some of the structure seen in the primary disk is the result of the recent close approach of ISO-Oph 2 B.
Stellar flybys are rare but still statistically possible 
in low-mass star-forming regions \citep{Cuello2020}.
Possible disk flybys reported in the literature include RW Aur \citep{Dai2015}, HV Tau and DO Tau \citep{Winter2018}, AS 205 \citep{Kurtovic2018} and UX Tau A \citep{Menard2020}. 
These flybys typically exhibit arcs of gas hundreds of au long connecting the disks in the system, in agreement with the results of hydrodynamic models (e.g. \citealt{Cuello2020}). 
The same models show that flybys also trigger spiral arms within the disks. Therefore,  we speculate that, \emph{if the initial configuration are two narrow rings}, a close approach could result in azimuthal  asymmetries similar to the ones observed in the disc around the ISO-Oph 2 A disk. 

Unfortunately the Gaia Data Release 2 only provides proper motions for the primary star and is not possible to constrain the relative motion of  ISO-Oph 2 A and ISO-Oph 2 B with enough precision to assess the likelihood of the flyby scenario.
However, we note that, assuming the closest encounter occurred 500 yr ago, as in the simulation presented by Cuello et el. (2020), the relative proper motion between the primary an the secondary would be expected to be $\sim$3.4 mas/yr. 
The flyby hypothesis could also be tested in the future with hydrodynamic modeling tailored to the system. 

The disk around ISO-Oph 2 B is also very interesting as it has one of the smallest cavities ever resolved at 
(sub)mm wavelengths.  With a radius of just $\sim$2.2 au, the cavity size is at the limit of ALMA's resolution at the distance of the closest molecular clouds (d $\sim$ 140 pc). 
Interestingly, TW Hydra, one of the closest stars to Earth with a protoplanetary disk  (d = 60 pc) also hosts a transition disk with a very small cavity, only $\sim$1 au in radius \citep{Andrews2016}. 
However, unlike TW Hydra, ISO-Oph 2 B has a disc that seems to be narrow (width/radius $<$ 1.0),  resembling the structures of other disks such as ISO-Oph 2 A, Sz 91 \citep{Canovas2016}, and T Cha \citep{Hendler2018}, but in a much smaller scale (by a factor of 15-40 in radius).    
Furthermore, ISO-Oph 2 B also is the lowest-mass object with a resolved dust cavity in the Ophiuchus molecular cloud, and possible in any region 
Based on the age ($\sim$1 Myr) and mass (0.08 $M_{\odot}$) estimates discussed in Section 3.1, ISO-Oph 2 B should have an M6.5 spectral type. However, all the 7 objects that have resolved cavities in the ODISEA sample have earlier spectral types (Ruiz-Rodriguez et al. in prep). %
That is also the case for all pre-main-sequence stars with transition disks identified in Ophiuchus based on their IR SEDs \citep{Cieza2010,Merin2010}; and all the transition objects with mm data (from all regions) recently compiled by \citet{Francis2020}, and \citet{Pinilla2020}. 

\section{Summary and Conclusions} 

We present new 1.3 mm ALMA long-baseline data of the ISO-Oph 2 system at 3 au resolution.  Our main results and conclusions are as follow:

1) The ring-like structure of the primary disk is resolved into two non-axysimmetric rings, suggesting dynamical disturbances.  Future hydrodynamic modeling could  help explaining the origin of these substructures and establish whether they are due to an internal or an external disturber.  

2) The high-resolution continuum observations reveal the disk around the secondary is 7 au across (FHWM) and has an inner cavity just 2.2 au in radius.
This implies that ISO-Oph 2 B has the smallest (transition) disk ever imaged in Ophiuchus and likely has a ring-like morphology (width $<$ inner radius).  
 
3) The flux ratio in the  of ISO-Oph 2 system at 2.2 $\mu$m (0.08) and the spectral type of the primary (M0) implies the secondary is at the brown dwarf mass limit, rendering ISO-Oph 2 B the lowest mass object in the cloud with a resolved dust cavity in its disk. 
 
4) The $^{12}$CO observations indicate that there is significant gas within the large cavity in the primary disk and suggest that the two disks might be connected by bridge of gas. This bridge might indicate that the secondary object recently flew by the primary and provides an alternative scenario for the origin of the azimuthal asymmetries in the disk around the primary.

\acknowledgments
We thank the anonymous referee for the helpful comments, and constructive remarks on this article.
This paper makes use of the following ALMA data: ADS/JAO.ALMA\#2016.1.00728.S and ADS/JAO.ALMA\#2018.1.00028.S.
ALMA is a partnership of ESO (representing its member states), NSF (USA) and
NINS (Japan), together with NRC (Canada) and NSC and ASIAA (Taiwan), in cooperation with the Republic of Chile. 
M.F. has received funding from the European Research Council (grant 757957).
A.Z. acknowledges support from the ANID-FONDECYT grant 11190837. 
S.P acknowledges support from ANID-FONDECYT grant 1191934 and the ESO-Chile Joint Committee. 
P.P. acknowledges support provided by the Alexander von Humboldt Foundation.
D.P. acknowledges support from Australian Research Council grants DP180104235 and FT130100034

\onecolumngrid

\begin{figure}
\gridline{\fig{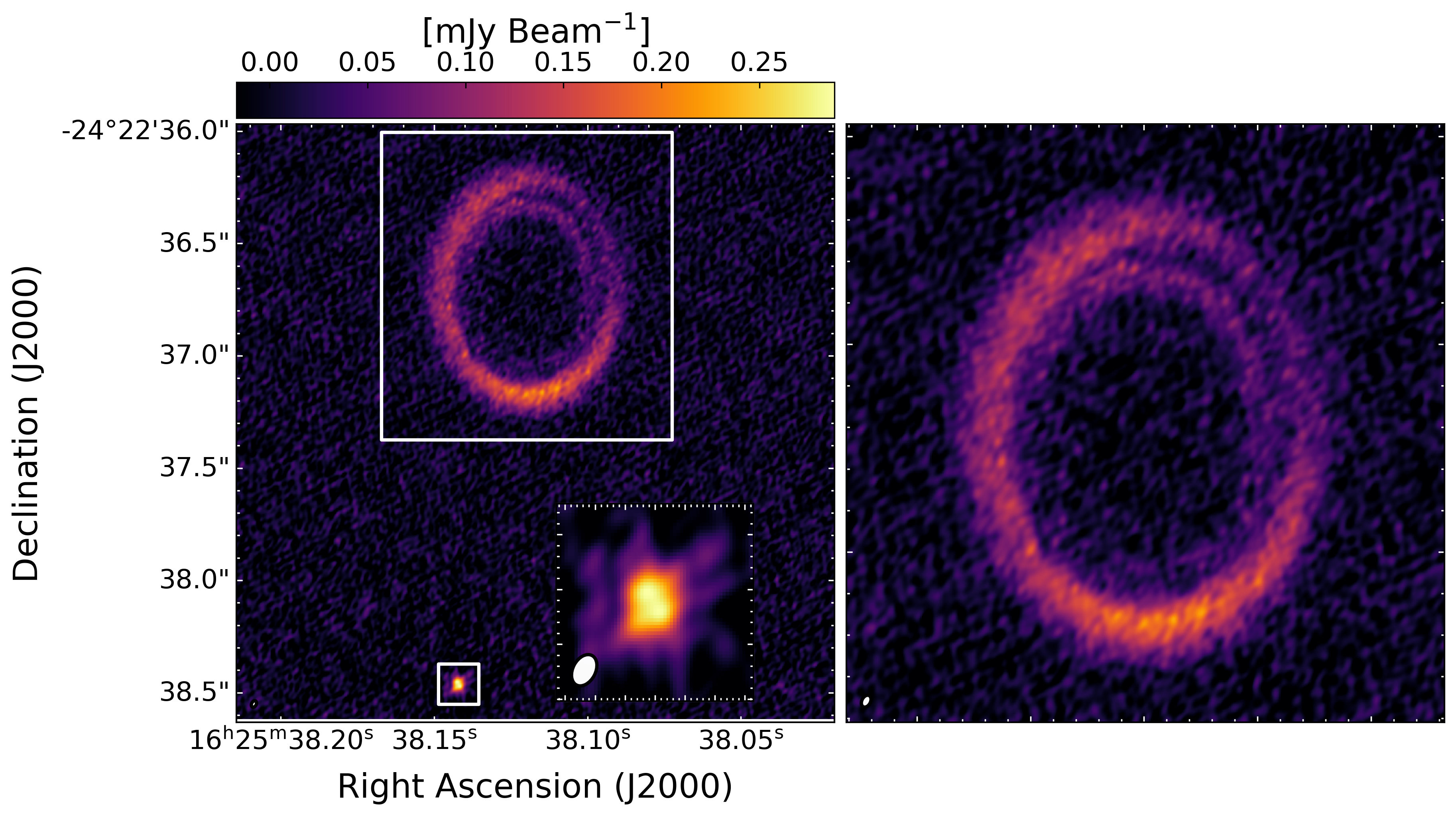}{0.9\textwidth}{}}
\caption{Left: 1.3 mm image of the ISO-Oph 2 system, including the disks around both stellar components and a zoom-in of the secondary disk where the cavity is marginally resolved. Right: A zoom-in of the primary disk with two non-axisymmetric rings. North is up and East is to the left.}
\end{figure}

\begin{figure}
\gridline{\fig{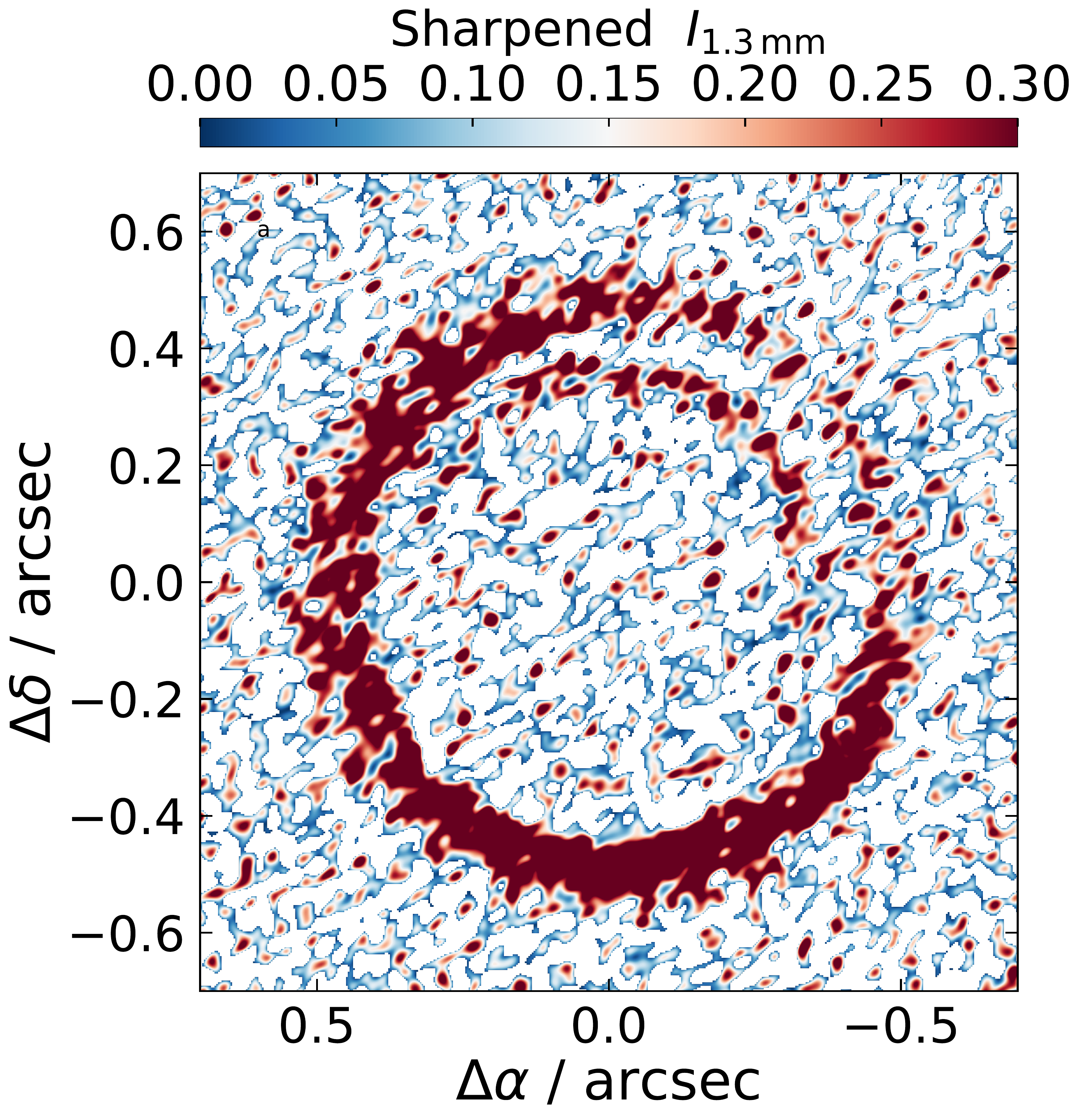}{0.43\textwidth}{(a)}
         \fig{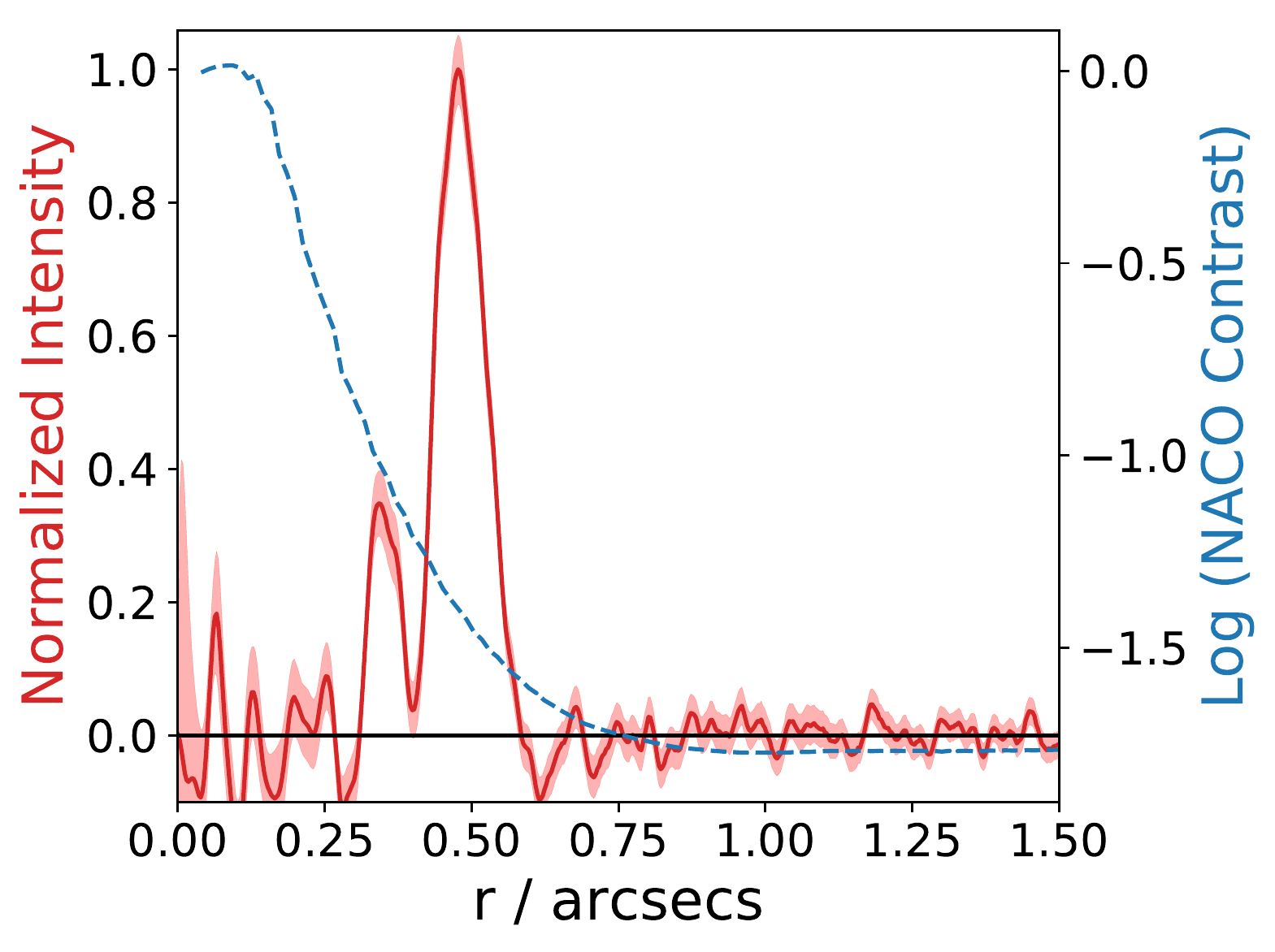}{0.53\textwidth}{(b)}}
\gridline{\fig{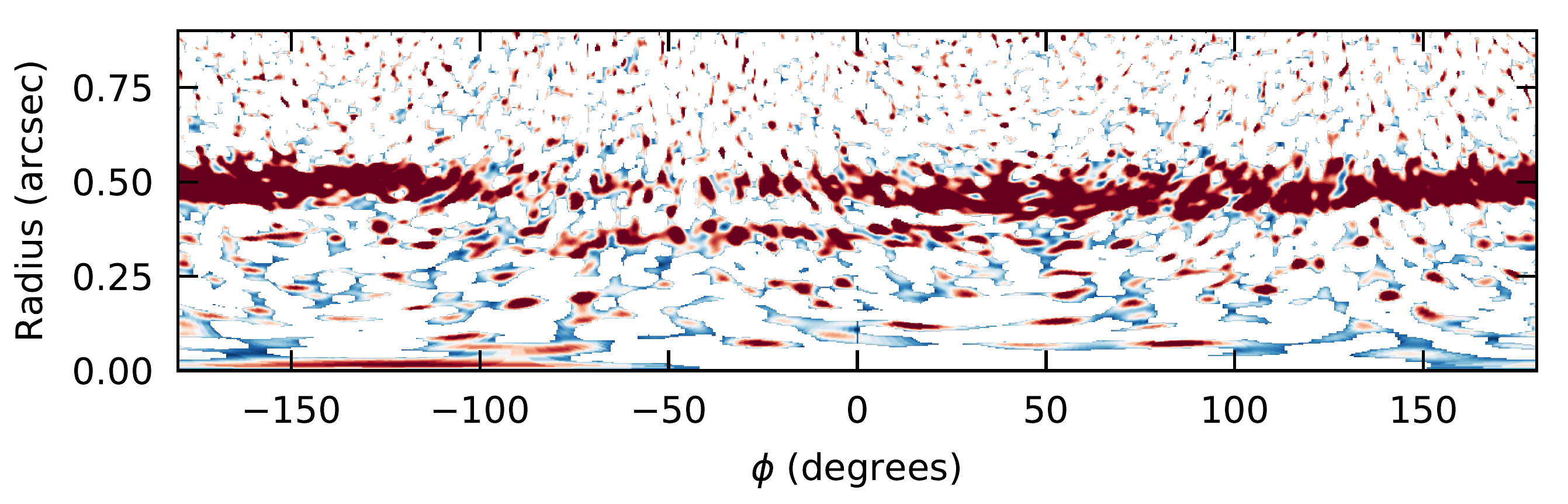}{0.9\textwidth}{(c)}}
\gridline{\fig{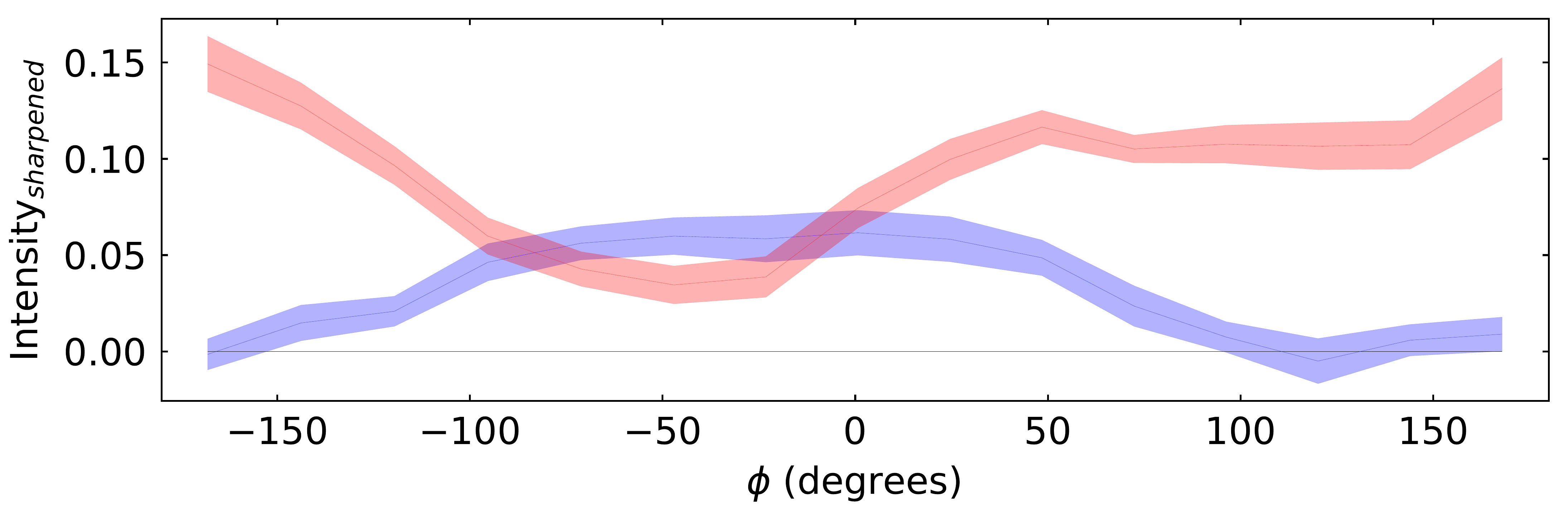}{0.9\textwidth}{(d)}}
\caption{(a) Deprojected image of the disk around the primary with normalized intensity to 0.3$\times$peak. (b) Average deprojected radial profile of the disk around the primary and the 5-$\sigma$ contrast curve of the VLT-NACO observations at 2.2 $\mu$m reported by \citet{Zurlo2020}. (c) Polar radial deprojection from image (a). (d) Same as panel (c), but averaged over radius, between 0.3-0.4$''$ and 0.4-0.6$''$, for the inner ring (blue line) and outer ring (red line), respectively. Shaded regions correspond to the errors in the mean.}  
\end{figure}

\begin{figure}
\gridline{\fig{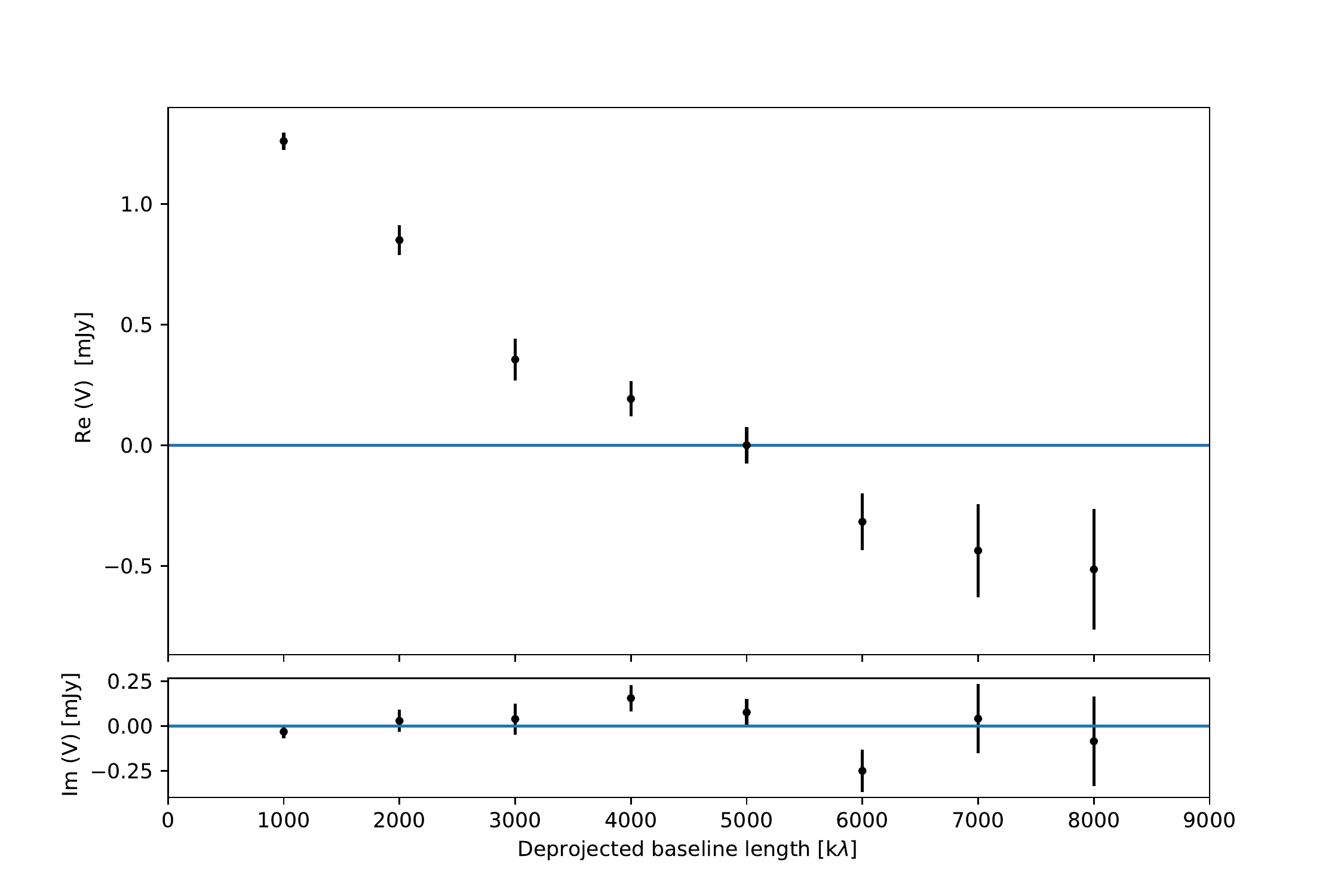}{0.9\textwidth}{(a)}}
\caption{The deprojected visibility profile of the disk around the secondary. Both the Real (top panel) and the Imaginary (bottom panel) parts are shown. A null is seen at a baseline length of $\sim$5000$\pm$500 k$\lambda$ (6500 m), indicating a dust cavity  with a radius of $\sim$2.2 AU.}
\end{figure}

\begin{figure}
\gridline{\fig{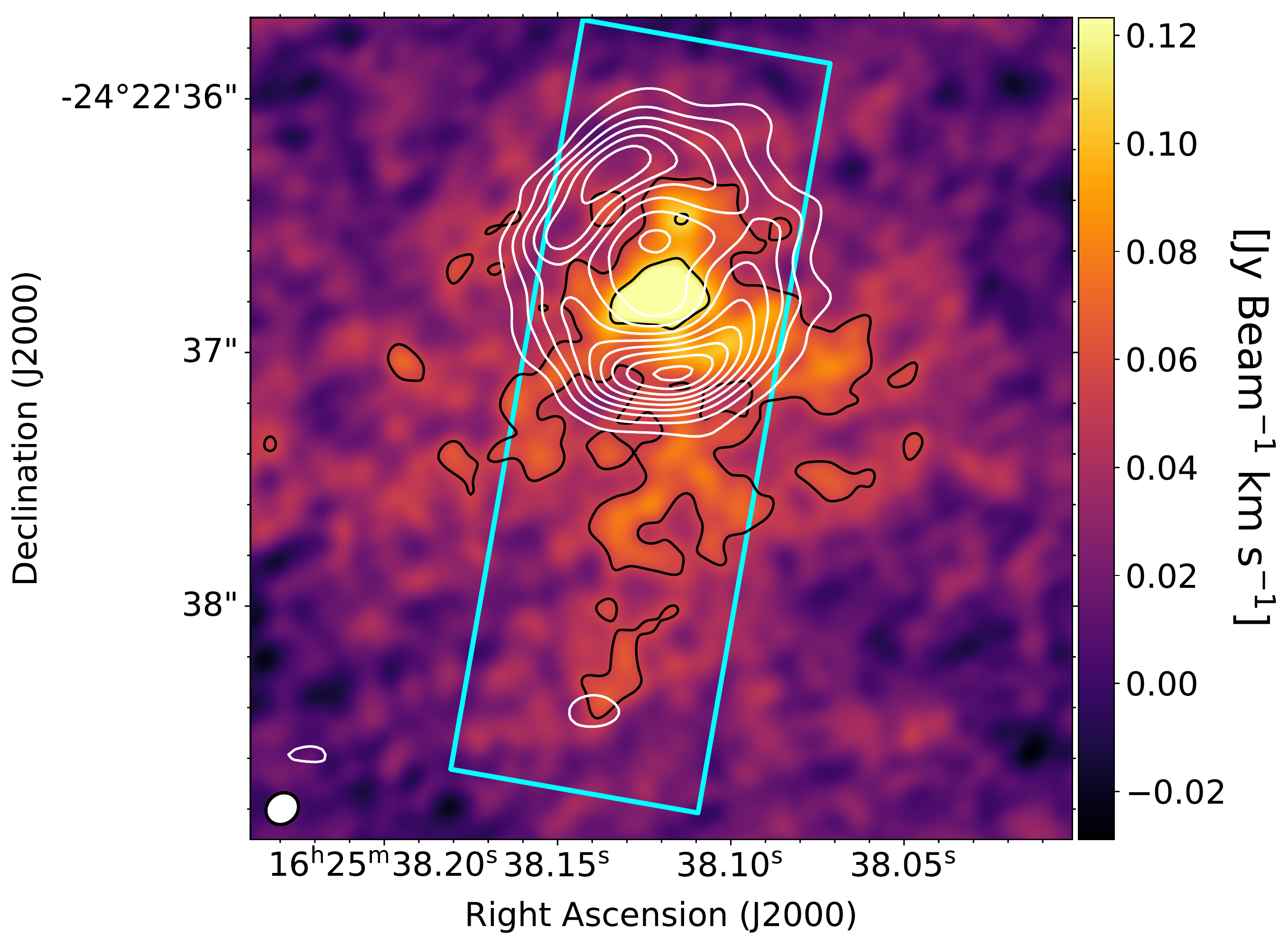}{0.58\textwidth}{(a)}
         \fig{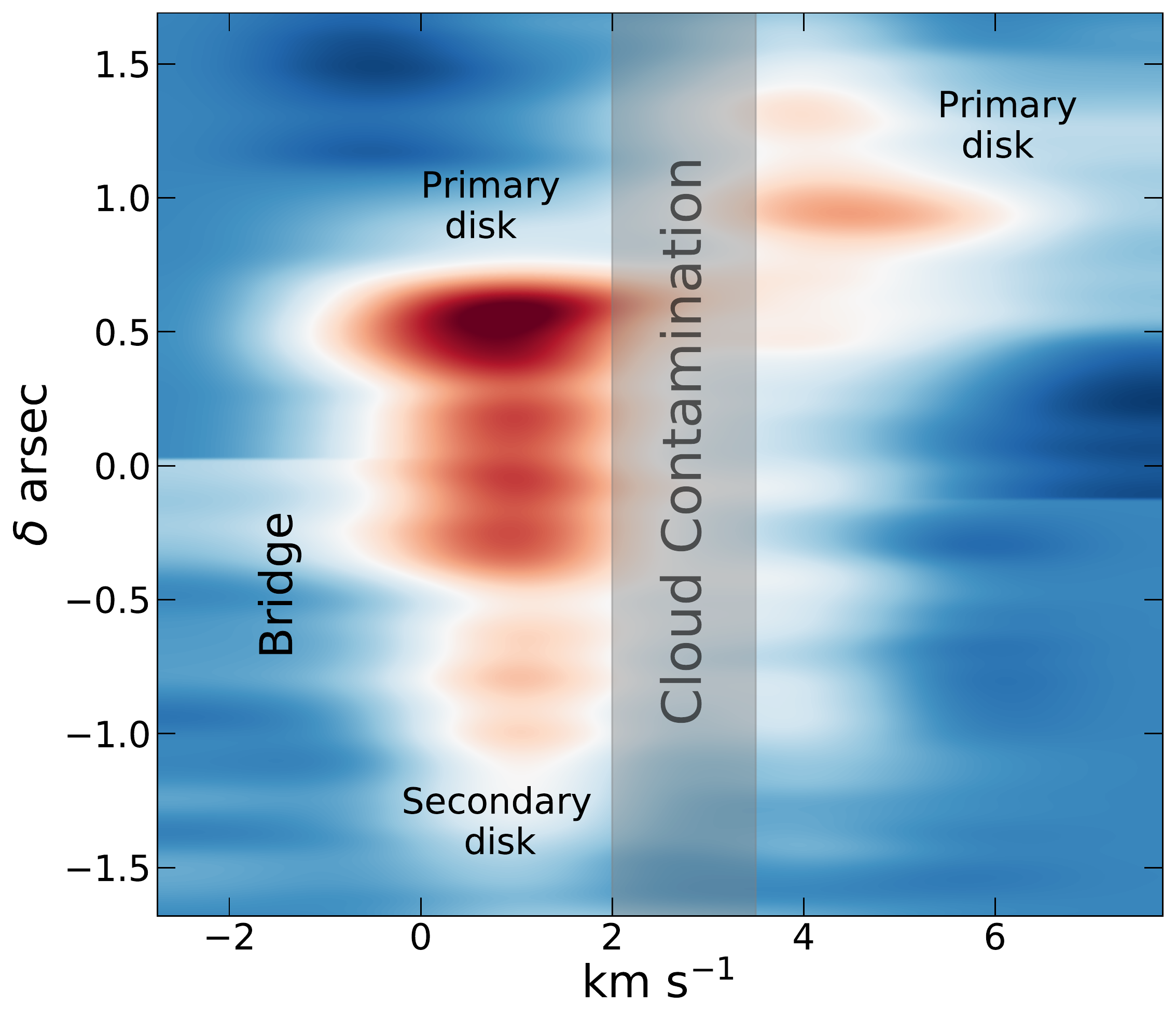}{0.49\textwidth}{(b)}}
\gridline{\fig{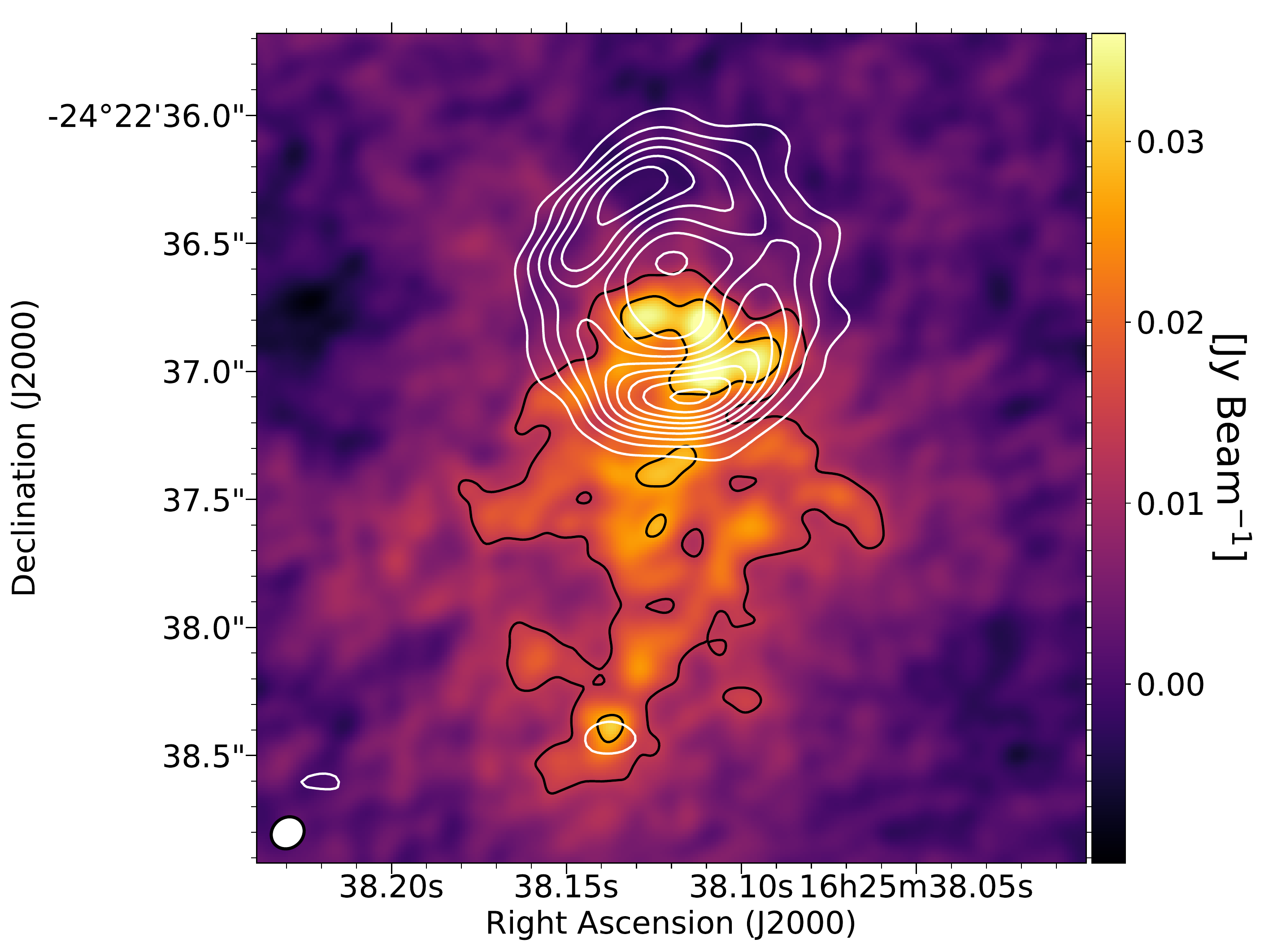}{0.55\textwidth}{(c)}
         \fig{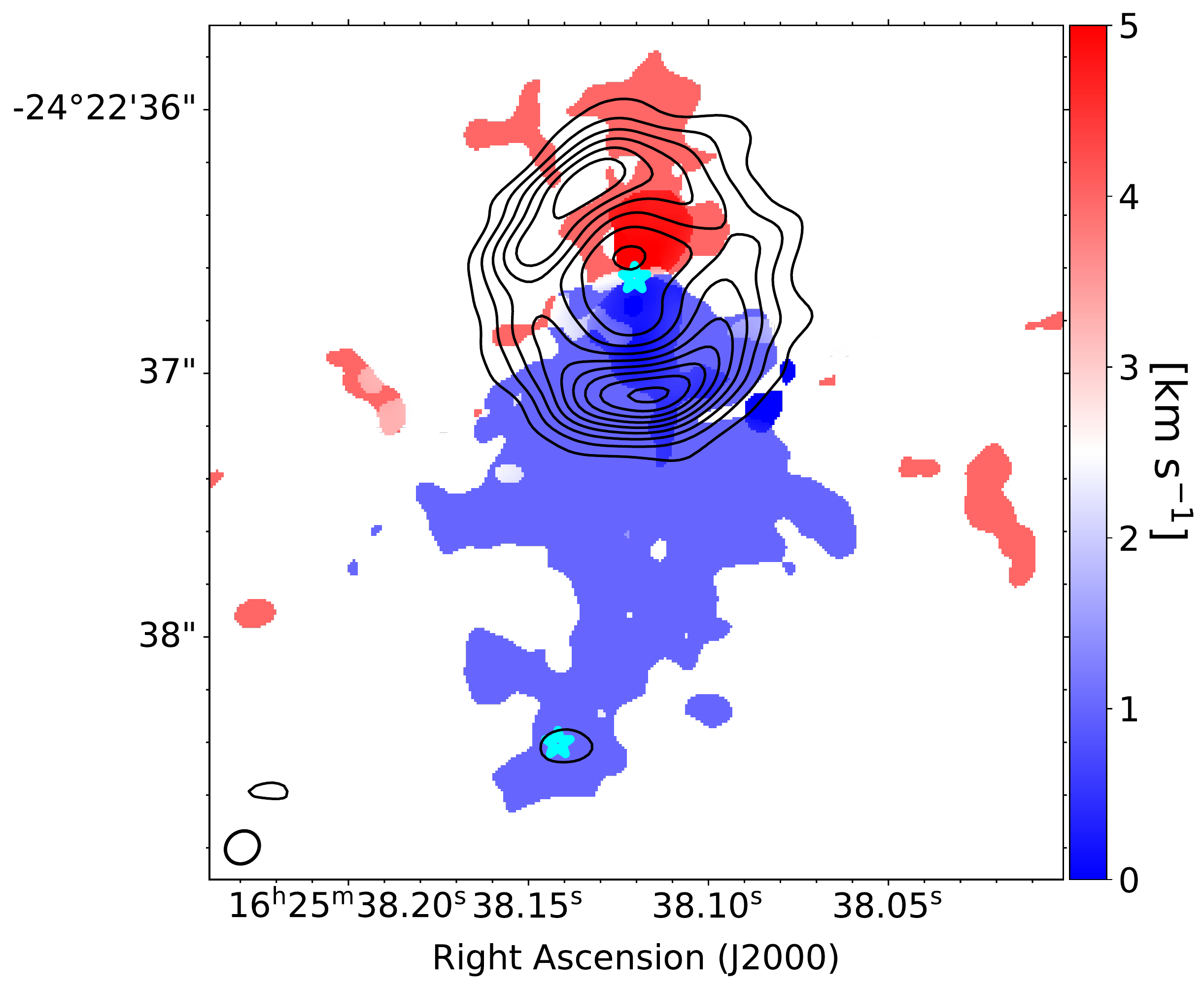}{0.51\textwidth}{(d)}}
\caption{Panel (a): the Moment-0 from the combined visibilities at 0.03$''$, 0.2$''$ and 1.1$''$ resolution imaged with an 0.1$''$ beam. The black contours represent emission at $\geq3-\sigma$ with steps of 3-$\sigma$. Continuum contours in 0.2$''$ are shown in white.
Panel (b): the Position-Velocity diagram  with $PA$=169$^{o}$, length=3.4$''$ and averaging width=1$''$ (see rectangle in panel a), revealing significant emission connecting both disks at $\sim$1.5 km/s.
Panel (c): the channel map corresponding to velocities
between 0.25 and 1.75 km/s.   Panel (d): the moment-1 of the same data from panel (a). The position of the stars from Gaia, corrected for the proper motion of the primary are also indicated.}   
\end{figure}

\newpage

\bibliography{sample63}{}
\bibliographystyle{aasjournal}



\end{document}